# Terahertz emission from ZnGeP$_2$: Phase-matching, intensity and length scalability

Joseph D. Rowley,[1] Derek A. Bas,[1] Kevin T. Zawilski,[2]

Peter G. Schunemann,[2] and Alan D. Bristow[1,*]

[1] Department of Physics, West Virginia University, Morgantown, West Virginia, 26506-6315, USA
[2] BAE Systems, MER15-1813 P.O. Box 868, Nashua, New Hampshire 03061, USA
*Corresponding author: alan.bristow@mail.wvu.edu

Collinear phase-matched optical rectification is studied in ZnGeP$_2$ pumped with near-infrared light. The pump-intensity dependence is presented for three crystal lengths (0.3, 1.0 and 3.0 mm) to determine the effects of linear optical absorption, nonlinear optical absorption and terahertz free-carrier absorption on the generation. Critical parameters such as the coherence length (for velocity matching), dispersion length (for linear pulse broadening) and nonlinear length (for self-phase modulation) are determined for this material. These parameters provide insight into the upper limit of pulse intensity and crystal length required to generate intense terahertz pulse without detriment to the pulse shape. It is found that for 1-mm thick ZnGeP$_2$(012), pumped at 1.28 μm with intensity of ~15 GW/cm$^2$ will produce intense undistorted pulses, whereas longer crystals or larger intensities modify the pulse shape to varying degrees. Moreover, phase-matching dispersion maps are presented for the terahertz generation over a large tuning range (1.1-2.4 μm) in longer (3 mm) crystal, demonstrating the phase-matching bandwidth and phase mismatch that leads to fringing associated with multi-pulse interference. All observed results are simulated numerically showing good qualitative agreement.

## 1. INTRODUCTION

Broadband terahertz (THz) pulses are of interest for spectroscopy in fundamental science [1,2] and commercial applications [3–6]. Optimal time- and frequency-domain attributes for the THz pulses vary significantly with each application. For example, intense electric-field amplitudes are important for nonlinear light-matter interactions [7,8], whereas a combination of broad bandwidth and stability are required for THz time-domain spectroscopy [9,10].

Optical rectification (OR) of femtosecond laser pulses has proven to be an efficient source of broadband THz pulses. To date, the highest conversion efficiencies have been achieved in LiNbO$_3$ with 0.8-μm pumping using the tilted-pulse-front geometry to achieve velocity matching between the pump and generated THz pulses [11]. Large tilt angles can introduce aberration, reduce bandwidth and diminish phase matching beyond ~2.5 THz. Moreover, LiNbO$_3$ is often pumped above the two-photon absorption (2PA) edge, leading to pump depletion and free-carrier absorption (FCA) of the THz radiation. These issues may be mitigated by using other materials [11,12], where pumping below the 2PA and even three-photon absorption (3PA) edge is possible, and by wide-area pumping in the collinear OR geometry, as demonstrated in ZnTe pumped at 0.8 μm [13,14].

GaP and GaAs are promising materials for collinear OR at infrared (IR) wavelengths longer than 0.8 μm [15–19], even at telecom wavelengths. These zincblende semiconductors exhibit 2PA at 1.05 μm and 1.75 μm and optimum collinear phase matching at 1.0 μm and 1.4 μm respectively. Phase matching is achieved in GaP at wavelengths longer than the 2PA edge, whereas single-crystal GaAs must be limited in thickness to achieve broadband phase-matching. Alternatively, quasi-phase matching in optically isotropic semiconductors can allow for thicker crystals, as demonstrated in GaP pumped at 1.55 μm [20] and in diffusion-bound or optically-contacted GaAs [21–23].

OR in chalcopyrite ZnGeP$_2$ (ZGP) shares attractive qualities with zincblende III-V semiconductors [24,25], including much lower FCA below ~3 THz compared with ZnTe and LiNbO$_3$, smaller velocity mismatch in the near IR (NIR), large 2$^{nd}$-order nonlinear coefficients, and a 2PA edge at 1.1 μm. Thus, 2PA can be avoided in ZGP at phase-matched pump wavelengths, as is the case in GaP. On the other hand, ZGP has a nonlinear coefficient approximately twice that of GaP and a nonlinear refractive index ~1/4 that of GaAs as summarized in [16]. These attributes make longer single-domain ZGP a good candidate for high-intensity pumping in the NIR. Moreover, the birefringence of ZGP allows natural tunability of the phase matching [25].

In this work the OR pump wavelength and intensity are considered for various ZGP crystal lengths to determine the envelope for high-intensity THz generation. Experimental data are presented and modeled with pump depletion, nonlinear optical absorption and FCA. Group-velocity dispersion (GVD) and self-phase modulation (SPM) are also discussed. In addition, pump wavelength-dependent measurements map the effect of velocity matching on the THz emission [26].

This paper is arranged as follows: In Section 2, a comparison of the coherence length is provided for GaP, GaAs and ZGP. In Section 3, experimental details are presented. In Section 4, results and discussion are presented for the intensity, crystal length and pump-wavelength measurements.

## 2. Comparison of the Coherence Length

The wavelength dependence of the group and phase velocities describes the propagation of optical and THz pulses through a crystal. The velocity-matching condition is characterized by the distance at which the group velocity of the pump pulse relative to the phase velocity of a particular THz frequency component have a π/2-phase difference. This coherence length is given as

$$L_c = \frac{c}{2\nu_{THz} \mid n_{THz} - n_g \mid}, \quad (1)$$

Where $c$ is the vacuum light speed $n_{THz}$ is the refractive index at the THz frequency $\nu_{THz}$, and $n_g = n - \lambda \partial n / \partial \lambda$ is the group index for pump carrier wavelength $\lambda$ and corresponding refractive index $n$.

The coherence length can be shown in a two-axis dispersion map. Figure 1 shows numerical calculations of the coherence length for GaAs(110), GaP(110), ZGP(110) and ZGP(012), based on literature values of the optical [27,28] and THz [29–31] dispersion relations. In ZGP the birefringence must be considered. Since maximum generation is achieved with polarization configuration *ooe* and *eeo* for the (110) and (012) crystals respectively [25], the latter dispersion values are modified to compensate for the pump polarization being at 45° to the z-axis of the index ellipsoid.

In all four cases shown in Fig.1 the range of the velocity-matched pump wavelengths increases toward lower $\nu_{THz}$ and that thicker crystals exhibit a narrower velocity matching range. Moreover, the optimal velocity-matched pump wavelength differs depending on crystal thickness. For example, in thick GaP crystals optimal velocity matching is achieved near 1 µm, but for thin crystals pumping near 0.75 µm potentially provides a wider THz bandwidth; see the horizontal lines in Fig. 1(b).

For GaAs and GaP velocity matching is disrupted by strong phonon absorption lines at 8.02 THz and 10.96 THz respectively. The former reduces the useful generation range in GaAs compared to GaP. In ZGP phonon absorption lines are expected at 11.03 THz for (110)-cut and 10.34 THz for (012)-cut crystals. Weaker phonons are also present at 3.60 THz, 4.26 THz, 6.06 THz, and 9.80 THz [31]. These weaker absorption lines will result in narrow absorption lines and disruption of phase matching near their respective frequencies. They are sufficiently weak to not prevent broadband THz generation from short excitation pulses. ZGP is well suited to excitation with 100-fs NIR/IR pulses to produce ~3-THz bandwidth pulses.

## 3. EXPERIMENTAL

Nominally 100-fs pulses from a 1-KHz Ti:sapphire laser amplifier and optical parametric amplifier (OPA) gate and generate THz pulses in a conventional time-domain emission spectroscopy setup [24]. The Ti:sapphire pulses have a center wavelength of 0.8 µm and the OPA is automatically tunable from 1.1 µm to 2.4 µm.

The OPA pulses impinge the sample at normal incidence (collinear geometry), at room temperature in a $N_2$-purge box to minimize water absorption. The pump beam is loosely focused to $1/e^2$ diameter of ~1.2 mm. The THz emission is collimated then refocused by a pair of identical off-axis parabolic mirrors. At the focal position of the second mirror electro-optic (EO) sampling is performed in a 0.5-mm ZnTe(110) crystal, gated by an 0.8-µm pulse. The gate pulse passes through a Soleil-Babinet compensator (set to ~λ/4) and a Wollaston prism onto balanced Si or Ge photodiodes. The balanced signal is measured in a lock-in amplifier referenced to a synched 300-Hz mechanical chopper. The linear response of the electro-optic sampling was confirmed by varying both the power of the gate with calibrated ND filters and the power of the THz using a series of up to five Si wafers, each with transmittance ~ 50% and resistivity ~ 10 MΩ.

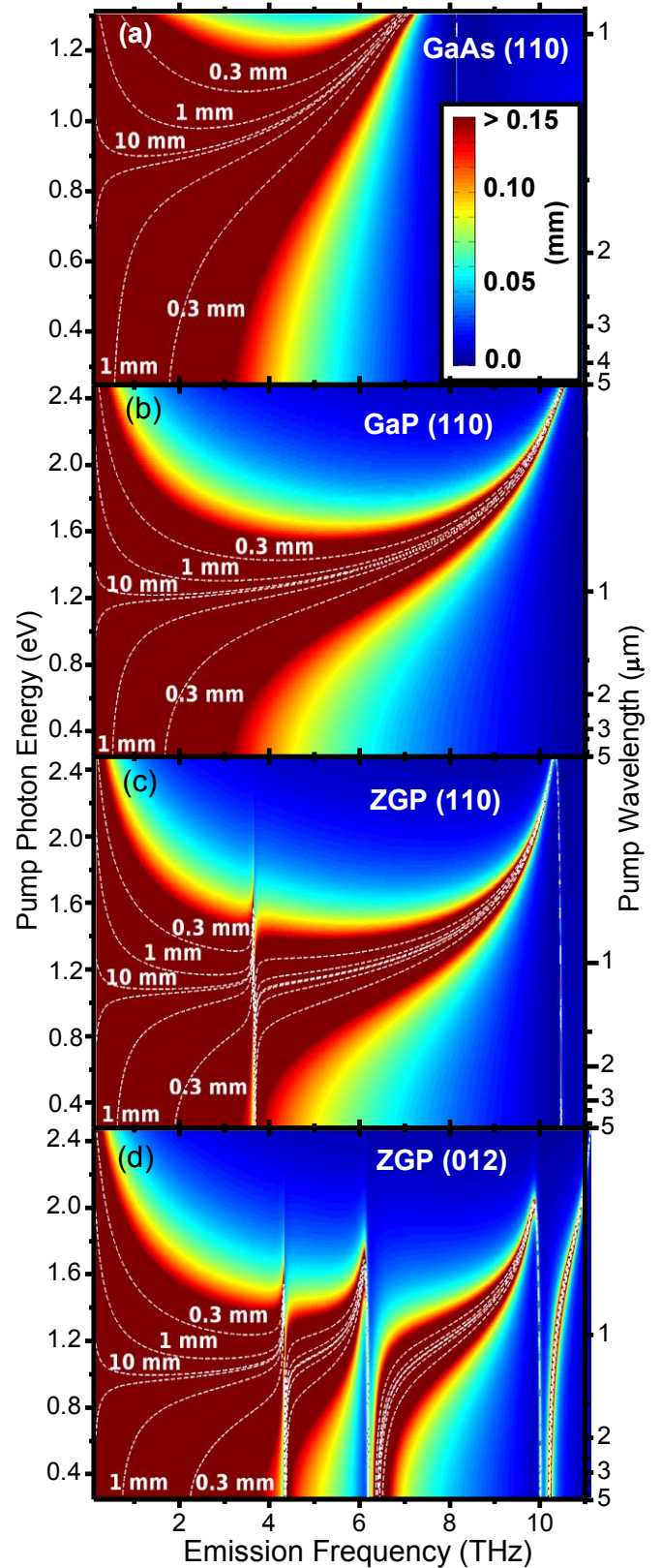

**Fig. 1** (Color online) Coherence length dispersion maps for (a) GaAs(110), (b) GaP (110), (c) ZnGeP2(110), and (d) ZnGeP2(012). The scale saturates at a coherence length of 0.15 mm. For 0.3 mm, 1 mm, and 10 mm thicknesses are enclosed within two dotted lines and labeled accordingly.

The source crystals, ZnTe detector, and the gate polarization are oriented for best generation and detection efficiency [25], with the pump pulse vertically polarized. Due to the tetragonal crystal structure and ~2% lattice compression in the ⟨001⟩ direction of ZGP the pump polarization leads to o-wave and e-wave excitation for the (110) and (012) crystals. Repeatability of the sample position is achieved by referencing to a fixed straight edge without altering the alignment of the setup.

Undoped single crystals of ZGP were grown from a stoichiometric melt by the horizontal gradient freeze technique [32] and double-side polished for optical measurements. Several samples were cut from the same boule: 0.33 mm, 0.92 mm, and 3.03 mm of ZGP(012) and 3.01 mm of ZGP(110).

## 4. RESULTS AND DISCUSSION

### A. Power and Length Scalability

Figure 2 (a) shows THz transients collected from the 0.33-mm, 0.92-mm and 3.03-mm ZGP(012) crystals, which are pumped with 1.28-μm radiation with intensity < 1 GW/cm². The pump wavelength and intensity ensure the best phase-matching condition and that nonlinear absorption is negligible [24]. ZGP(012) is studied, because ZGP(110) is phase matched at the edge of the OPA tuning range; see Section 4B.

The THz pulse shapes are 2nd-order derivate in accordance with OR. The first feature $E_1$ is negative due to the phase of the lock-in amplifier, the second feature $E_2$ is positive and the third feature $E_3$ is also negative. Features at later times can be ignored, since they occur through limitations of the detection crystal bandwidth and water absorption of the THz radiation. Transients are normalized to $E_1$, which allows for the identification of pulse shape changes in the low power regime. As the crystal thickness increases $E_1$ and $E_2$ appear to be unaffected, but there is a reduction in the magnitude of $E_3$, accompanied by a change in time delay ($\Delta t_{13}=t_3-t_1$). Previously, shape changes of this sort have been attributed to FCA by photoinjected carriers and occur at higher intensities [33], but are not due to shift currents [34].

Figure 2 (b) shows THz transients collected from the 0.33-mm, 0.92-mm and 3.03-mm crystals at pump intensity 35 GW/cm². A low-intensity transient for the 0.92-mm crystal is shown for reference. Transients are normalized to $E_1$. At high intensity the pulse distortion is more pronounced than at lower powers, which may be attributed to several nonlinear optical effects.

Figure 2(c) shows the spectral response of both extremes of the intensity and crystal thickness ranges, namely the 0.33-mm crystal pumped at <1 GW/cm² and the 3.03-mm crystal pumped at 35 GW/cm². The spectra are obtained from a numerical Fourier transform of the transients and then normalized to their respective peaks. For the thin sample spectral components extend to near 3.5 THz. There is strong bandwidth reduction for the thick crystal pumped at high intensity, indicating FCA of the high-frequency THz components and/or pump-pulse stretching.

Figures 3(a) - (c) show the extracted magnitudes of the $E_1$, $E_2$ and $E_3$ features for the three crystal thicknesses over a power range from ~1 GW/cm² to 35 GW/cm². In all three features the magnitude increases linearly for low pump intensities, with

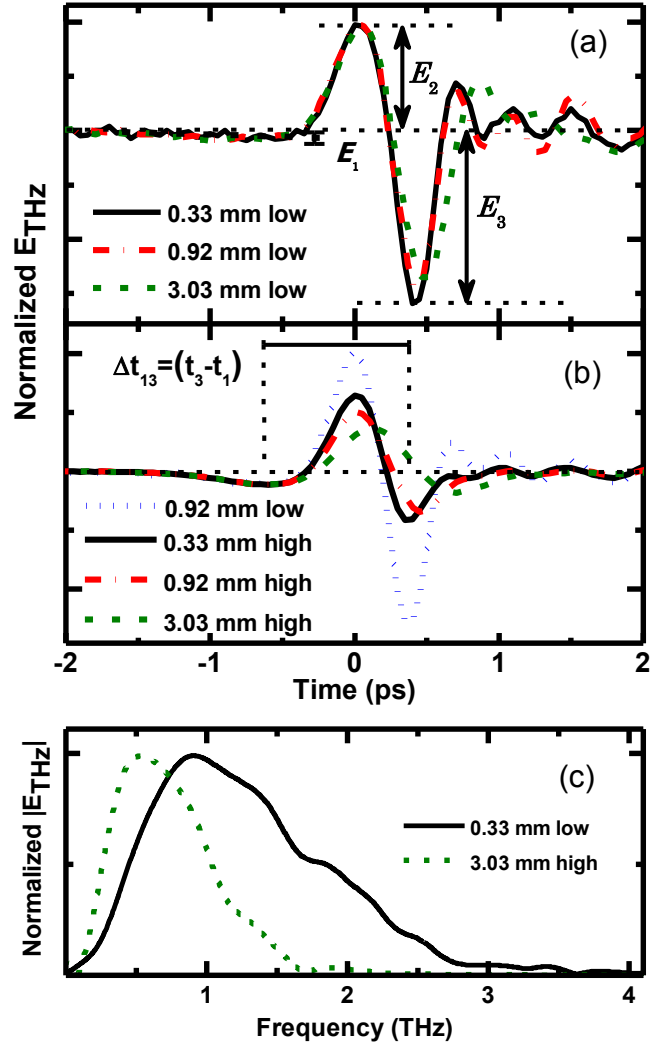

**Fig. 2** (color online) Terahertz transient from 0.33-mm, 0.92-mm and 3.03-mm thick ZGP(012) at (a) low and (b) high pump intensity. (c) Spectra for the two extremes of the intensity and thickness ranges.

varying degrees of sublinear response at higher intensities. Assuming perfect phase matching, the linear response of the THz electric-field is $\propto d_{ijk}I$, where $d_{ijk}$ is the appropriate 2nd-order nonlinear coefficient and $I$ is the pump intensity. The pump intensity is governed by the linear and nonlinear absorption in accordance with Beer's law

$$\frac{\partial I(z,t)}{\partial z} = -\alpha_{opt}I(z,t) - \gamma I^3(z,t), \qquad (2)$$

where $z$ is the position inside the crystal, $t$ is time, $\alpha_{opt}$ is the linear optical absorption and $\gamma$ is the 3PA coefficient. 2PA is neglected due to selected excitation wavelength [24].

The NIR linear and nonlinear absorption coefficients are determined from intensity-dependent pump-depletion measurements. Figure 3(d) shows the pump-depletion results with intensity inside the first interface on the horizontal axis and the transmission intensity ($I_{Trans}$) on the vertical axis. All aspects of this pump-depletion experiment were identical to the pumping conditions for the data shown in Fig. 3(a)-(c). For all three crystals the transmission is reduced at low intensity due to reflection and linear absorption. With increasing intensity, the transmission for the thinner crystals increases nearly linearly, but for the thicker crystal strong sublinear behavior is observed.

Solving equation (2) for a Gaussian pulse while accounting for reflectivity from both surfaces gives the transmittance [35,36]

$$T = \frac{I_{Trans}}{I_0} = \frac{(1-R)^2 e^{-\alpha_{opt} L}}{\sqrt{\pi} q_o} \int_{-\infty}^{\infty} \ln\left[\sqrt{1+q_o^2 e^{-2t^2}} + q_o e^{-t^2}\right] dt \quad (3)$$

where $R$ is the reflectance calculated from the index of refraction, $L$ is the crystal length and the Gaussian beam parameter is

$$q_o = [2\gamma(1-R)^2 I_o^2 [1-\exp(-2\alpha_{opt} L)] / 2\alpha_{opt}]^{1/2}, \quad (4)$$

where $I_0$ is the incident intensity. Fitting the pump-depletion data with equation (3) and (4) yields values of $\alpha_{opt}$ = 1.07 cm$^{-1}$ and $\gamma$ = 0.01 cm$^3$/GW$^2$.

The THz intensity-dependence can be modeled with the following expression

$$\frac{\partial E_{THz}(z,t)}{\partial z} \propto C d_{eff} I(z,t) - E_{THz}(z,t) \times \left[\frac{\alpha_{THz}}{2} + \alpha_{FCA}(t)\left(\alpha_{opt} I(z,t) + \frac{\gamma}{3} I^3(z,t)\right)\right], \quad (5)$$

which taking into account the generation due to the effective 2$^{nd}$-order nonlinear coefficient $d_{eff}$ = 39.4 pm/V and the measured linear optical absorption $\alpha_{opt}$, nonlinear optical absorption $\gamma$ and linear THz absorption $\alpha_{THz}$ = 3.15 cm$^{-1}$. The latter is obtained from the THz transmission through the various length samples (data not shown). The remaining free parameters are the FCA absorption coefficient $\alpha_{FCA}$ and adjustable scaling parameter $C$ for the purpose of modeling $E_1$, $E_2$ and $E_3$ individually, since the strength of each feature is inherently different and the exact scaling of the experimental data is arbitrary. Table 1 shows values of $C$ and $\alpha_{FCA}$ determined from numerically solving Equation (5) for all three crystal thicknesses simultaneously.

Table 1. Modeling Parameters for Intensity-Dependent THz Field Amplitude Features Extracted from the Transients for Various ZGP(012) Crystal Thicknesses

|  | $C$(Arb. Units) | | | $\alpha_{FCA}(t)$ |
|---|---|---|---|---|
|  | 0.33 mm | 0.92 mm | 3.03 mm | (cm$^2$/GW) |
| $E_1$ | 22.5 | 22.5 / 1.17 | 22.5 / 1.41 | 0 |
| $E_2$ | 190 | 190 / 1.17 | 190 / 1.41 | 0 |
| $E_3$ | 260 | 260 / 1.17 | 260 / 1.41 | 0.33 |

The simulations match the experimental data extremely well for $E_1$ and $E_2$ in the thinner crystal, where it is expected that FCA is negligible [33]. Indeed FCA is necessary to model $E_3$, since the trailing edge of the pulse occurs significantly later in time and is affected by the photoinjected carriers. This means that linear, nonlinear and free-carrier absorption are almost sufficient to explain the intensity dependence of the thinner ZGP.

Overall, the generation of THz radiation increases with crystal thickness and the intensity dependences exhibit an increase in their slope; as seen in the data. However, the increased slope decreases for longer crystals, i.e. the increase in slope between the thin and medium crystals is larger than that between the medium and thick crystals. This effect is also highlighted by the multipliers 1/1.117 and 1/1.41 for the 0.92 mm and 3.03 mm crystals respectively. The consistent decrease in the scaling factor cannot be attributed to any of the absorption terms described above. Instead, it may be related to the pulse spreading due to GVD, as characterized by the linear dispersion length

$$L_D = \tau_o^2 \bigg/ \left(\frac{d^2k}{d\omega^2}\right)_{\omega_o} \quad (6)$$

where $\tau_o$ is the initial pump pulse duration, $k$ is the wave vector, $\omega$ is the angular frequency and $\omega_0$ is the angular carrier frequency. Excitation of ZGP by a transform-limited Gaussian pump pulse with center wavelength of 1280 nm and pulse width of 120 fs gives $L_D$ = 12 mm. A propagating transform-limited pulse doubles in temporal width at 0.6 $L_D$. Suggesting that for

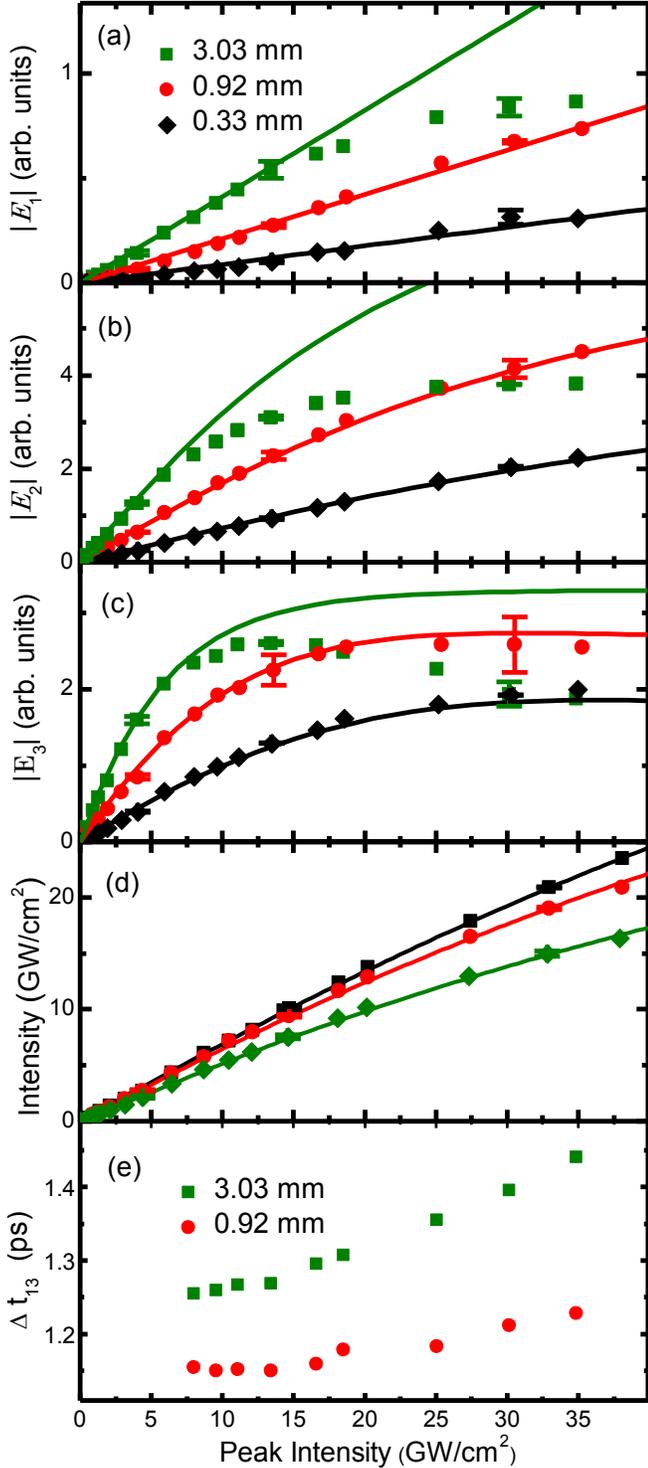

Fig. 3 (a) – (c) (Color online) Pump intensity terahertz electric field amplitude for the $E_1$, $E_2$ and $E_3$ features for three crystal thicknesses with, numerical simulations. (d) Pump pulse depletion data and best fit. (e) Time delay between $E_1$ and $E_3$ features for the thicker ZGP crystals.

the 3-mm thick ZGP the excitation pulse has been stretched by ~25%, which is non-negligible and is expected to be the cause in the decrease of the slope with increasing crystal length.

For the thickest crystal, data in Fig. 3(a)-(c) departs from the model at high pump intensity for the $E_1$, $E_2$ and $E_3$ features, which can only be due to an additional nonlinear process, such as SPM of the pump pulse. SPM stretches the pump pulse in time and results in stretching of the THz pulse, as seen in Fig. 3(e) and reduction of the THz spectral bandwidth as seen in Fig. 2(c). SPM is verified with measurements of the spectral width after the ZGP crystal were performed, showing a ~350% increase over the full range of pump intensities (data not shown), but without significant loss of the Gaussian shape.

SPM is due to a nonlinear change in refraction $n_2 I_0(z,t)$ across the pulse in both space and time. If the resulting nonlinear phase shift $\varphi_{NL}(z,t)$ at $\omega_0$ is ≥1 radian then this will be observed in the pulse shape. The accumulation of SPM can be defined by a characteristic nonlinear length in the propagation direction [35]

$$L_{NL} = \frac{c}{\omega_o n_2 I_o}, \quad (7)$$

where $n_2$ = 40 cm²/GW for ZGP [16]. For $I_0$ = 40 GW/cm² then $L_{NL}$ = 1.3 mm, which suggests that SPM will be a limiting factor in the efficiency of THz generation for thick crystals at high intensity pumping.

## B. Mapping the Effects of Phase-Matching

Figure 4(a) shows the maximum THz field amplitude, defined as $E_{PP} = |E_2| + |E_3|$, at low intensity (1.5 GW/cm²) over the tuning range of the OPA signal (1.1 µm – 1.62 µm) in 20 nm intervals for the 3-mm thick ZGP(110) and ZGP(012) crystals. Each crystal cut was oriented to achieve maximum near-infrared to THz conversion efficiency. The wavelength of peak conversion efficiency are ~1.1 µm for ZGP(110) and ~1.3 µm for ZGP(012). These values agree well with the best coherence lengths shown in Fig. 1 and confirm previously results [24,25].

To better understand the consequences of phase mismatching in the time and frequency domain measurements are performed over the entire tuning range of the OPA (1.1 µm – 2.5 µm). Figure 4(b) shows the temporal response for ZGP(110). For phase-matched THz generation at ~1.1 µm a single 2nd-order differential pulse is observed, identical to that seen in Fig. 2(a). Increasing the pump wavelength away from the phase-matching condition changes the observed THz waveforms and reduces the overall signal strength. All transients are therefore arbitrarily normalized to the strongest $E_2$ feature (dashed line) to emphasize the evolution of the waveforms.

Large phase mismatch exhibits splitting of the THz pulse into pulses, traveling at the pump group velocity $v_g$ and THz phase velocity $v_{THz}$ [37,38]. As the pump wavelength increases $v_g$ also increases, hence the leading pulse arrives at the detector earlier. However, tuning the OPA introduces small artificial time delays into the data. The delay can be removed by shifting each transient so that $E_2$ (or the leading peak) approximately follows the calculated relative change in group time delay, (dashed line). As expected the trailing pulse, traveling at THz phase velocity, remains constant in time for sufficiently large phase mismatch.

Figs. 4(c) and (d) show the spectral response of the ZGP(110) and ZGP(012) crystals respectively, obtained by a numerical Fourier transform. Each individual spectrum is normalized to the highest peak. A broad continuous spectrum only occurs where phase-matching is achieved. The observed phase-matching

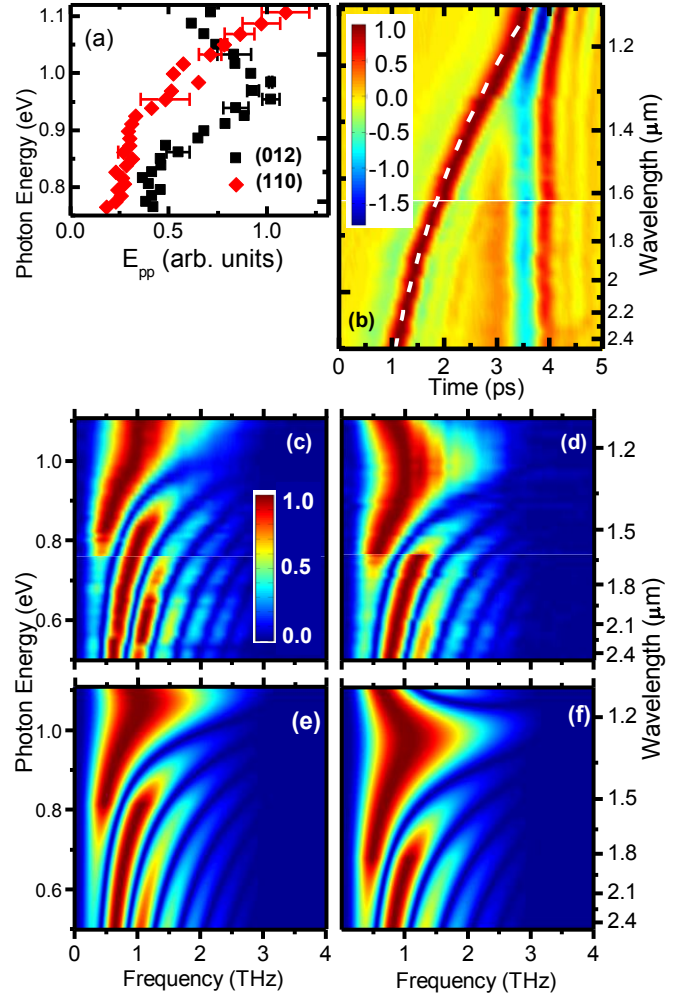

**Fig. 4** (color online) Terahertz emission for (012)- and (110)-cut ZGP crystals as a function of pump wavelength. (a) The extracted $E_{pp}$, (b) the full temporal response for (110) including of the calculated relative group delay time (dashed line), (c) Fourier transform of the temporal data for (110), (d) Fourier transform of the temporal data for (012), (e) simulated (110) response and (f) simulated (012) response.

bandwidth for pumping ZGP(012) is ~80 nm, which is centered at 1280 nm. While the THz bandwidth remains approximately constant with pump tuning, fringes begin to appear with a spectral period that increases as $v_g$ and $v_{THz}$ diverge. This occurs because the two pulses arriving at the detector acquire increasing time delay that introduces a greater spectral phase [39].

The entire spectral response can be express as the superposition of three terms [37]

$$E_{THz}(z,\omega) = R^{(2)} E_o^2 [-e^{(i\omega n_g z/c)} + \tfrac{1}{2}(1+n_R)e^{i\omega n_{THz} z/c} + \tfrac{1}{2}(1-n_R)e^{(-i\omega n_{THz} z/c)}, \quad (8)$$

where $R^{(2)} = 2\sqrt{2}\pi d_{eff}(\omega)\tau_0 e^{-\omega^2 \tau_0^2/4} / (n_{THz}^2 - n_g^2)$ is the nonlinear generation function and $n_R = n_g / n_{THz}$. The first term represents a forward propagating wave at the group velocity of the pump pulse, the second term is a forward propagating wave at a THz phase velocity and third term is a negligible backward propagating wave at a THz phase velocity. Due to interference in the bulk of the crystal, THz emission from only the surfaces are radiated toward the detector, hence two weaker pulses are observed in the temporal data. Finally, to account for the frequency-dependent diffraction and propagation loss, as well as

EO-detector bandwidth limitations, transfer functions [$T_{diff}(\omega)$, $T_{focus}(\omega)$, $T_{det}(\omega)$] can be applied to Equation (8), as prescribed by Faure *et al.* [37].

Figure 4 (e) and (f) show the simulated spectral response for the ZGP(110) and ZGP(012) crystals respectively, based on Equation (8) and appropriate transfer functions. Very good qualitative agreement is achieved between experiment and the model, which reproduces the fringing, the broadband phase matched regions and the drop in detected signal at low THz frequencies due to diffraction of the THz pulse at the source. Input parameters for the model include the measured crystal thicknesses of 3.01 mm (110) and 3.03 µm (012), pump pulse FWHM duration of 100 fs, $1/r^2$ pump radius of 0.85 mm, probe wavelength of 800 nm, detection crystal thickness of 0.5 mm and probe pulse duration of 300 fs.

This dispersion mapping was performed with the 3-mm thick crystal to emphasis the fringing away from the phase-matching regime. For thinner crystals the phase-matched pulse remains intact over a much larger range of pump wavelengths as is predicted by the coherence length calculations.

## CONCLUSIONS

This work comprises an extensive study of ZnGeP$_2$ as a collinear optical rectification source of pulsed terahertz radiation. Overall, ZnGeP$_2$ appears to be a superior source compared to GaP or GaAs pumped in the infrared. Hence, moderate length crystals are useful for pumping by modelocked fiber laser in the telecom band.

There are three critical lengths that should be considered in the design of a terahertz source: First, the coherence length $L_c$, which defines the useful frequency range for velocity matching; second, the dispersion length $L_D$, which defines the linear-dispersion-induced broadening of the pump pulse; and third, the nonlinear length $L_{NL}$, which defines the nonlinear distortion of the pump pulse by self-phase modulation. In addition to these critical lengths, the single and multiphoton absorption coefficients for the pump and the free-carrier absorption for the terahertz must be considered. From these considerations the useful envelope of the THz generation from ZnGeP$_2$ is determined.

This work also demonstrates that for moderate length ZnGeP$_2$ crystals (~1 mm) pumped with high intensity (~15 GW/cm$^2$) the terahertz pulse shape and bandwidth are maintained, making this crystal suitable for wide-area excitation to produce intense pulses of terahertz radiation in the collinear pump geometry.

## ACKNOWLEDGEMENTS

The work at WVU was partly sponsored by the West Virginia Higher Education Policy Commission (HEPC.dsr.12.29) and the National Science Foundation (CBET-1233795). JDR wishes to thank the WVNano Initiative for support.

## REFERENCES

1. R. Ulbricht, E. Hendry, J. Shan, T. F. Heinz, and M. Bonn, "Carrier dynamics in semiconductors studied with time-resolved terahertz spectroscopy," Rev. Mod. Phys. **83**, 543–586 (2011).
2. T. Kampfrath, A. Sell, G. Klatt, A. Pashkin, S. Mährlein, T. Dekorsy, M. Wolf, M. Fiebig, A. Leitenstorfer, and R. Huber, "Coherent terahertz control of antiferromagnetic spin waves," Nat. Photon. **5**, 31–34 (2011).
3. Z. D. Taylor, R. S. Singh, D. B. Bennett, P. Tewari, C. P. Kealey, N. Bajwa, M. O. Culjat, A. Stojadinovic, H. Lee, J.-P. Hubschman, E. R. Brown, and W. S. Grundfest, "THz medical imaging: in vivo hydration sensing," T. THz Sci. Technol., IEEE **1**, 201 –219 (2011).
4. J. Federici and L. Moeller, "Review of terahertz and subterahertz wireless communications," J. Appl. Phys. **107**, 111101 (2010).
5. M. Tonouchi, "Cutting-edge terahertz technology," Nat. Photon. **1**, 97–105 (2007).
6. H.-W. Hübers, "Terahertz technology: Towards THz integrated photonics," Nat. Photon. **4**, 503–504 (2010).
7. F. Blanchard, G. Sharma, L. Razzari, X. Ropagnol, H.-C. Bandulet, F. Vidal, R. Morandotti, J.-C. Kieffer, T. Ozaki, H. Tiedje, H. Haugen, M. Reid, and F. Hegmann, "Generation of intense terahertz radiation via optical methods," J. S. Top. Quantum Electron., IEEE **17**, 5 –16 (2011).
8. Matthias C Hoffmann and József András Fülöp, "Intense ultrashort terahertz pulses: generation and applications," J. Phys. D: Appl. Phys. **44**, 1–13 (2011).
9. M. Hangyo, M. Tani, and T. Nagashima, "Terahertz time-domain spectroscopy of solids: a review," Int. J. Infrared Milli. **26**, 1661–1690 (2005).
10. C. A. Schmuttenmaer, "Exploring dynamics in the far-infrared with terahertz spectroscopy," Chem. Rev. **104**, 1759–1780 (2004).
11. J. A. Fülöp, L. Pálfalvi, G. Almási, and J. Hebling, "Design of high-energy terahertz sources based on optical rectification," Opt. Express **18**, 12311–12327 (2010).
12. M. I. Bakunov, S. B. Bodrov, and E. A. Mashkovich, "Terahertz generation with tilted-front laser pulses: dynamic theory for low-absorbing crystals," J. Opt. Soc. Am. B **28**, 1724-1734 (2011).
13. T. Löffler, T. Hahn, M. Thomson, F. Jacob, and H. Roskos, "Large-area electro-optic ZnTe terahertz emitters," Opt. Express **13**, 5353–5362 (2005).
14. F. Blanchard, L. Razzari, H. C. Bandulet, G. Sharma, R. Morandotti, J. C. Kieffer, T. Ozaki, M. Reid, H. F. Tiedje, H. K. Haugen, and F. A. Hegmann, "Generation of 1.5 µJ single-cycle terahertz pulses by optical rectification from a large aperture ZnTe crystal," Opt. Express **15**, 13212–13220 (2007).
15. M. C. Hoffmann, K.-L. Yeh, J. Hebling, and K. A. Nelson, "Efficient terahertz generation by optical rectification at 1035 nm," Opt. Express **15**, 11706–11713 (2007).
16. K. l. Vodopyanov, "Optical THz-wave generation with periodically-inverted GaAs," Laser Photonics Rev. **2**, 11–25 (2008).
17. J.-P. Negel, R. Hegenbarth, A. Steinmann, B. Metzger, F. Hoos, and H. Giessen, "Compact and cost-effective scheme for THz generation via optical rectification in GaP and GaAs using novel fs laser oscillators," Appl. Phys. B **103**, 45–50 (2011).

18. M. Nagai, K. Tanaka, H. Ohtake, T. Bessho, T. Sugiura, T. Hirosumi, and M. Yoshida, "Generation and detection of terahertz radiation by electro-optical process in GaAs using 1.56 µm fiber laser pulses," Appl. Phys. Lett. **85**, 3974–3976 (2004).
19. G. Chang, C. J. Divin, C.-H. Liu, S. L. Williamson, A. Galvanauskas, and T. B. Norris, "Power scalable compact THz system based on an ultrafast Yb-doped fiber amplifier," Opt. Express **14**, 7909-7913 (2006).
20. I. Tomita, H. Suzuki, H. Ito, H. Takenouchi, K. Ajito, R. Rungsawang, and Y. Ueno, "Terahertz-wave generation from quasi-phase-matched GaP for 1.55 µm pumping," Appl. Phys. Lett. **88**, 071118 (2006).
21. L. Gordon, G. L. Woods, R. C. Eckardt, R. R. Route, R. S. Feigelson, M. M. Fejer, and R. Byer, "Diffusion-bonded stacked GaAs for quasiphase-matched second-harmonic generation of a carbon dioxide laser," Electron. Lett. **29**, 1942–1944 (1993).
22. L. A. Eyres, P. J. Tourreau, T. J. Pinguet, C. B. Ebert, J. S. Harris, M. M. Fejer, L. Becouarn, B. Gerard, and E. Lallier, "All-epitaxial fabrication of thick, orientation-patterned GaAs films for nonlinear optical frequency conversion," Appl. Phys. Lett. **79**, 904–906 (2001).
23. Y.-S. Lee, W. C. Hurlbut, K. L. Vodopyanov, M. M. Fejer, and V. G. Kozlov, "Generation of multicycle terahertz pulses via optical rectification in periodically inverted GaAs structures," Appl. Phys. Lett. **89**, 181104 (2006).
24. J. D. Rowley, J. K. Pierce, A. T. Brant, L. E. Halliburton, N. C. Giles, P. G. Schunemann, and A. D. Bristow, "Broadband terahertz pulse emission from $ZnGeP_2$," Opt. Lett. **37**, 788–790 (2012).
25. J. D. Rowley, J. K. Wahlstrand, K. T. Zawilski, P. G. Schunemann, N. C. Giles, and A. D. Bristow, "Terahertz generation by optical rectification in uniaxial birefringent crystals," Opt. Express **20**, 16968–16973 (2012).
26. N. C. J. van der Valk, P. C. M. Planken, A. N. Buijserd, and H. J. Bakker, "Influence of pump wavelength and crystal length on the phase matching of optical rectification," J. Opt. Soc. Am. B **22**, 1714–1718 (2005).
27. T. Skauli, P. S. Kuo, K. L. Vodopyanov, T. J. Pinguet, O. Levi, L. A. Eyres, J. S. Harris, M. M. Fejer, B. Gerard, L. Becouarn, and E. Lallier, "Improved dispersion relations for GaAs and applications to nonlinear optics," J. Appl. Phys. **94**, 6447–6455 (2003).
28. F. L. Madarasz, J. O. Dimmock, N. Dietz, and K. J. Bachmann, "Sellmeier parameters for $ZnGaP_2$ and GaP," J. Appl. Phys. **87**, 1564–1565 (2000).
29. W. J. Moore and R. T. Holm, "Infrared dielectric constant of gallium arsenide," J. Appl. Phys. **80**, 6939–6942 (1996).
30. A. S. Barker, "Dielectric dispersion and phonon line shape in gallium phosphide," Phys. Rev. **165**, 917–922 (1968).
31. V. V. Voitsekhovskii, A. A. Volkov, G. A. Komandin, and Y. A. Shakir, "Dielectric properties of $ZnGeP_2$ in the far infrared," Phys. Sol. Stat. **37**, 1198–1199 (1995).
32. K. T. Zawilski, P. G. Schunemann, S. D. Setzler, and T. M. Pollak, "Large aperture single crystal $ZnGeP_2$ for high-energy applications," J. Crystal Growth **310**, 1891–1896 (2008).
33. S. M. Harrel, R. L. Milot, J. M. Schleicher, and C. A. Schmuttenmaer, "Influence of free-carrier absorption on terahertz generation from ZnTe(110)," J. Appl. Phys. **107**, 033526 (2010).
34. D. Côté, N. Laman, and H. M. van Driel, "Rectification and shift currents in GaAs," Appl. Phys. Lett. **80**, 905–907 (2002).
35. R. L. Sutherland, *Handbook of Nonlinear Optics* (CRC Press, 2003).
36. K. V. Adarsh, K. S. Sangunni, C. S. S. Sandeep, R. Philip, S. Kokenyesi, and V. Takats, "Observation of three-photon absorption and saturation of two-photon absorption in amorphous nanolayered $Se/As_2S_3$ thin film structures," J. Appl. Phys. **102**, 026102 (2007).
37. J. Faure, J. V. Tilborg, R. A. Kaindl, and W. P. Leemans, "Modelling laser-based table-top THz sources: optical rectification, propagation and electro-optic sampling," Opt. Quant. Electron. **36**, 681–697 (2004).
38. K. Wynne and J. J. Carey, "An integrated description of terahertz generation through optical rectification, charge transfer, and current surge," Opt. Comm. **256**, 400–413 (2005).
39. L. Lepetit, G. Cheriaux, and M. Joffre, "Linear techniques of phase measurement by femtosecond spectral interferometry for applications in spectroscopy," J. Opt. Soc. Am. B **12**, 2467–2474 (1995).